\documentclass[twocolumn]{aastex701}

\usepackage[utf8]{inputenc}
\usepackage[T1]{fontenc}

\begin{document}

\title{Limb Shift of the Fe I 6569 \AA\ line on the Sun }

\author[orcid=0009-0005-8311-1703]{H. C. Yu}
\affiliation{School of Astronomy and Space Science, Nanjing University, 163 Xianlin Road, Nanjing 210023, PR China}
\affiliation{Key Laboratory for Modern Astronomy and Astrophysics, Nanjing 210023, 163 Xianlin Road, PR China}
\email{DZ21260012@smail.nju.edu.cn}

\author[]{M. D. Ding} 
\affiliation{School of Astronomy and Space Science, Nanjing University, 163 Xianlin Road, Nanjing 210023, PR China}
\affiliation{Key Laboratory for Modern Astronomy and Astrophysics, Nanjing 210023, 163 Xianlin Road, PR China}
\email{dmd@nju.edu.cn}

\author[]{J. Hong} 
\affiliation{Institute for Solar Physics, Dept. of Astronomy, Stockholm University, AlbaNova University Centre, SE-106 91 Stockholm, Sweden}
\email{jie.hong@astro.su.se}

\author[]{Y. K. Wang} 
\affiliation{School of Astronomy and Space Science, Nanjing University, 163 Xianlin Road, Nanjing 210023, PR China}
\affiliation{Key Laboratory for Modern Astronomy and Astrophysics, Nanjing 210023, 163 Xianlin Road, PR China}
\email{wyk@nju.edu.cn}

\author[]{Z. Li}
\affiliation{School of Astronomy and Space Science, Nanjing University, 163 Xianlin Road, Nanjing 210023, PR China}
\affiliation{Key Laboratory for Modern Astronomy and Astrophysics, Nanjing 210023, 163 Xianlin Road, PR China}
\email{lizhen@nju.edu.cn}

\begin{abstract}
The convective motions of solar granules generate a center-to-limb variation of Doppler velocity in the photospheric lines, known as the limb shift effect. This study presents a comprehensive analysis of this effect for the Fe I 6569 \AA\ line using both observational data from the CHASE mission and numerical simulations from the Bifrost code. We employ two independent methods to derive the limb shift curve: a spectral-averaging method (Method 1) and a velocity-averaging method (Method 2). By comparing synthetic and observed data, we determine the convective blueshift, which is not accounted for in the CHASE observations. The simulations reproduce the observed trends for both methods at the instrument’s spatial resolution ($=1.2^{\prime\prime}$). However, at resolutions below $<1^{\prime\prime}$, Method 2 produces limb-shift curves that depart significantly from both Method-1 results and traditional limb-shift profiles, whereas Method 1 remains in agreement with classical behavior. Further analysis finds that the results from Method 1 comprise two distinct components: a contrast contribution caused by the correlation between velocity and line depth, and a Dopplergram contribution caused by density inhomogeneities and corrugation effects.

\end{abstract}


\section{Introduction}
Dopplergram of the photospheric lines of the solar disk shows an apparent redshift near the disk limb, which is the well-known limb shift. After subtracting other global velocity fields like differential rotation and meridional circulation, the residual velocity patterns are centrosymmetric, i.e., simply related to the heliocentric angle $\theta$. Such a relationship has been measured by sufficient works \citep{1960Higgs,1988Ulrich,2018Bottcher}. The order of limb shift is about 0.5 km/s, and the specific value depends on line depth and the lower excitation potential \citep{1981Dravins,2000Asplund}.

Researchers have proved that the limb shifts are mainly caused by the granular motion of the photosphere \citep{1967Bray,1978Beckers(a),1978Beckers(b)}. Granules are hot, upward-moving plasma cells surrounded by the cold, downward-moving intergranular network. When computing the spatially averaged spectrum, the contribution of bright blueshifted granules dominates over that of dark redshifted intergranular lanes due to their higher intensity, resulting in a net blueshift on the solar disk center \citep{1975Dravins}. However, these effects are not obvious at the solar limb, where the projection effect reduces the apparent radial velocity. As a result, the limb spectrum exhibits a relatively redshift compared to the disk center. In addition, the horizontal motion of the granules also contributes to the limb shift curve, producing an initial blueshift enhancement at intermediate heliocentric angles before the redshift emerges toward the limb \citep{1978Beckers(b)}.

Following the definition of limb shift \citep{1976Beckers}, the average velocity is supposed to be measured as \textbf{Spectral-Averaging Method} (Method 1 hereafter): Spatially average spectral line profiles at each heliocentric angle and then calculate the Doppler velocity using space-averaged line profiles \citep{1978Beckers(b),1982Dravins,2011Rodriguez,2018Bottcher,2018Cegla,2023Pietrow}. Method 1 contains the inhomogeneity of brightness by averaging the spectrum, so the brighter parts have a greater weight. 

At the same time, there is another common method adopted by plenty of works, which is \textbf{Velocity-Averaging Method} (Method 2 hereafter): First, calculate the Doppler velocity of each pixel to obtain the Dopplergram, then average the velocity across different heliocentric angles $\mathrm{\theta}$ \citep{1988Ulrich,2010Ulrich,2021Kashyap,2024Rao}. Method 2 inherently filters out the inhomogeneity, because each pixel has the same weight in the Dopplergram. 

It seems that different methods lead to similar results (see Fig.4 in \cite{2010Ulrich} and Fig.10 in \cite{2018Bottcher}). One possible reason is the pixel convolution effect. At low spatial resolution, the spectrum of each pixel contains the inhomogeneous brightness information of the granulation pattern.  However, how the spatial resolution influences the difference between these two methods needs further research. In addition, the spectral resolution also has an impact on the limb shift \citep{2011Rodriguez,2018Bottcher}.

Numerical simulations have reproduced the granular pattern successfully (e.g., \cite{1998Stein,2000Asplund,2007Cheung,2016Carlsson}). Several of them have computed the limb shift profile, demonstrating good agreement with observational measurements \citep{1978Beckers(b),1985Cavallini,2018Bottcher}. Moreover, \cite{2011Rodriguez} synthesizes the spectrum of neutral iron to calibrate the wavelength. Using simulation data, we are able to calculate the limb shift curve through two distinct approaches mentioned above. 

In this work, we investigate the limb shift of the Fe I 6569.2 \AA. The line asymmetry effect, such as the C-shape bisector, is not included \citep{2011Rodriguez}. The observation data come from the Chinese H$\alpha$ Solar Explorer (CHASE). Section~\ref{CHASE_Observation} provides a detailed description of the CHASE instrument and specifies Method 1 and Method 2. In Sect~\ref{Simulation}, we synthesize the Fe I spectrum based on the Bifrost simulation data. In Sect~\ref{Result}, we show results of the limb shift curve and analyze its dependence on spatial and spectral resolution. A discussion of physical differences between the two approaches is demonstrated in Sect~\ref{Origin}. After that, we discuss the application of these two methods in Sect~\ref{Discussion}, followed by a conclusion in Sect~\ref{conclusion}.

\section{CHASE Observation} \label{CHASE_Observation}
\begin{figure*}[htbp]
\centering
\includegraphics[width=\textwidth]{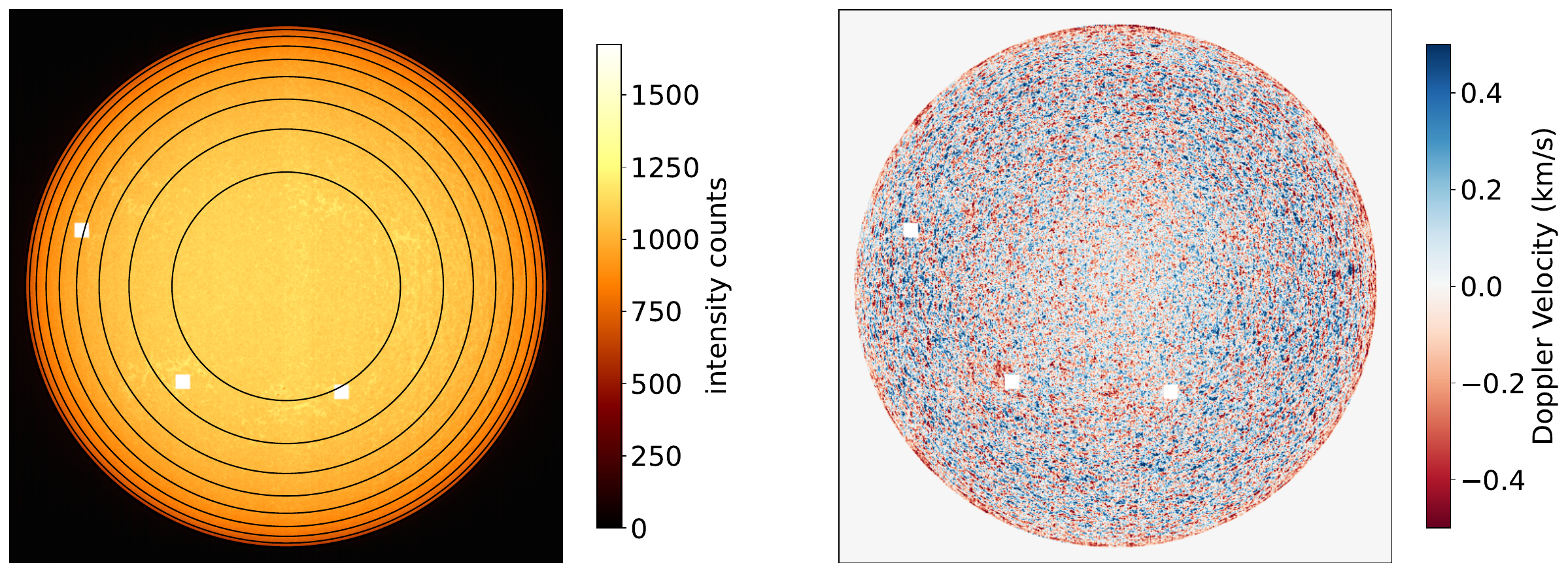}
\caption{Left: solar image in the Fe I line center. The concentric circles from center to limb represent the $\mu$ varying from 0.9 to 0, which divide the disk into 10 parts. Right: Dopplergram of the same line derived by bisectors; the color denotes the blueshift and redshift. Three white rectangles in both panels are the sunspot regions deduced when averaging.}
\label{solar_disk}
\end{figure*}

The CHASE acquires seeing-free spectroscopic observations by scanning the whole solar disk at two bands of H$\alpha$ (6562.8 \AA) and Fe I (6569.2 \AA), with a spectral sampling of 0.024 \AA\ \citep{2019CHASE}. Each scan takes 46 s, which produces an image with 1.2 arcsecond spatial resolution and 1 minute temporal resolution. We select data from one of the scans on August 9th, 2022. The Fe I (6569.2 \AA) band is chosen to study the photospheric information of limb shift. Usually, a binning of data is used to reduce the data size. Here, we use the data without binning for a higher spectral and spatial resolution. The data is composed of 4608 $\times $ 4625 pixels, and each pixel has 87 wavelength samples. The whole disk is then divided into 9 concentric rings based on heliocentric angle, from $\mu=0.1$ to 1 with a step of 0.1, where $\mu = \mathrm{\cos\theta}$ hereafter (see Figure~\ref{solar_disk}). The image is slightly shifted at different wavelengths due to the instrument's errors, which becomes serious when it comes too close to the limb. Therefore, the most limb region ($\mu < 0.1$) is not discussed in this work. 

We use method 1 and method 2 to calculate the limb shift curve. Method 1: First, average the pixels on different rings at each wavelength and then calculate the Doppler velocity using space-averaged line profiles. Method 2: First, calculate the Doppler velocity at each pixel to obtain the Dopplergram, then average the velocity in different rings. The mean value of the ring between $\mu_1$ and $\mu_2$ is represented by an effective $\bar{\mu}$, calculated as
\begin{equation}
    \bar{\mu}=\frac{\int^{\mu_2}_{\mu_1}\mu^2\mathrm{d}\mu}{\int^{\mu_2}_{\mu_1}\mu\mathrm{d}\mu}.
\end{equation}
The scale of a granule is roughly 1 Mm, thus each of the rings contains at least 100000 granules, which greatly reduces the influence of the 5-minute p-mode oscillation (As discussed in \cite{2011Rodriguez}). The effect of supergranulation is also reduced by averaging across the ring \citep{2023Ellwarth}. In addition to limb shift, the averaged doppler shift of on the solar disk is influenced by the following factors, (a) solar rotation, (b) meridional circulation, (c) fake doppler signal caused by relative motion of the satellite and the Sun, (d) gravitational redshift, (e) high magnetic field region such as sunspots \citep{2018Cegla}, and the impact of them is supposed to be eliminated. 

In Method 2, we first subtract the satellite motion from the Dopplergram to remove its contribution. Then we fit the Dopplergram of the solar disk with the sum of the solar rotation, meridional circulation, and limb shift. The fitting function of the solar rotation and meridional circulation is the same as \cite{2021Kashyap}. For the limb shift, \cite{2024Palumbo} uses a constant value with $\mu$ to retain the trend after subtracting; here we aim to obtain the limb shift curve with zero velocity at the disk center, so this constant is set to 0. The Gravitational redshift is a constant value across the whole disk, and it is eliminated by setting the velocity relative to the disk center. For the activity areas, we identify three sunspot regions and manually deduct these parts when averaging (Figure~\ref{solar_disk}). We then sum the fitted components, subtract them from the Dopplergram, and apply Method 2 to the resulting image (right panel of Fig~\ref{solar_disk}).

When applying Method 1, we have to preprocess the data before space averaging. The fitted image of Method 2 is eliminated from the raw spectrum by shifting the spectrum of each pixel. Consequently, the spectral lines contain only wavelength shifts caused by local motions (refer to \cite{2018Bottcher}).

We use a bisector to derive the Doppler velocity of the spectrum in Method 2 or the averaged spectrum in Method 1. The spectrum is first linearly interpolated to a finer wavelength grid. Then we select $I_0$ which makes $I_{\lambda_\mathrm{red}}=I_{\lambda_\mathrm{blue}}=I_0$. The doppler velocity can be calculated as $v=[\lambda_0-(\lambda_{\mathrm{red}}+\lambda_{\mathrm{blue}})/2]\cdot c/\lambda_0$, where $c$ is the speed of light and $\lambda_0$ is the wavelength in the line center. As is well known, the velocity is expected to depend on the value of $I_0$ \citep{1981Dravins}, indicating the line asymmetry. \citet{2011Rodriguez,2018Bottcher} calculate the velocity of different bisector intensities. For most photospheric lines (including Fe I 6569 \AA\ in our test), the bisector profiles feature a pronounced C-shape, with the maximum displacement occurring near the 0.7 continuum level. To extract the most representative value while reducing computational cost, we choose $I_0 = I_{\lambda_0}+0.7\cdot (I_{\lambda_0}-I_{\lambda_c})$ for each spectrum, where $\lambda_c=\lambda_0+0.5$ \AA.

\begin{figure}[htbp!]
\includegraphics[width=\linewidth]{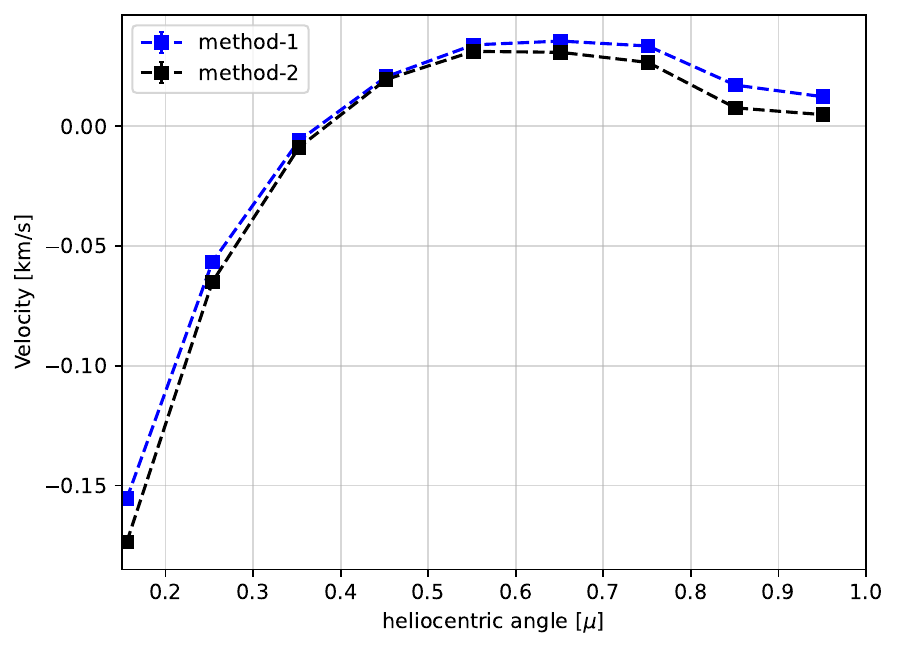}
\caption{Limb shift curves and their error bar as functions of the cosine of the heliocentric angle calculated with two methods. Positive value denotes blueshift. Velocity at the extreme limb is not shown.}
\label{limbshift_curve}
\end{figure}

The limb shift curve derived from two different methods is shown in Figure~\ref{limbshift_curve}. The two curves are nearly the same, but the intensity of the second curve is slightly lower. From center to limb, the blue-shift velocity gradually increases and reaches the top value when $\mu$ is about 0.65, then it changes rapidly to the red shift. All the values are the relative velocity with respect to the disk center because the Doppler velocity at the disk center is regarded as zero, and we will discuss it in Sec~\ref{Calibration}. In addition, we calculate the results with the data of four hemispheres (shown in Fig~\ref{hemisphere}), and we do not find significant discrepancies among different hemispheres.

\section{Simulation} \label{Simulation}

\subsection{Bifrost Data}
For comparison with the observation, we use numerical simulation data to synthesize the Fe I 6569.2 \AA\ line. The Bifrost is a state-of-the-art 3-dimensional radiative MHD model, which adds specific radiative transfer in the MHD system of equations \citep{2011Gudiksen}. Some of the simulation data is available 
\footnote{\url{https://sdc.uio.no/search/simulations}}. In this work, we use simulation en024048$\_$hion. The 3D box is $24\times24\times17$ $\mathrm{Mm}^3$ with 24 km horizontal resolution ($0.066^{\prime\prime}$) and 19-100 km vertical resolution. The box extends from 2.4 Mm below the visible surface to the corona. The convective motion is well reproduced, and more than 400 granules are included.

In the simulation, the energy of p-modes is spread over a very limited set of modes, giving them much larger amplitudes compared with the real Sun. The overall atmosphere oscillates at about 1 km/s at the photosphere with a period of 420 s \citep{2016Carlsson}. We follow \cite{1988Ulrich}, who defined the limb shift to be generated from matter having a zero net average mass flux along the line-of-sight. Therefore, data are selected at t = 5280 s, which has almost zero net mass flux in the vertical direction at the lower photosphere. Since the line forms at about 250 km above the $\tau_{500}=1$ \citep{2022Hong}, we only consider the layer from -200 km to 800 km. All properties are linearly interpolated at 10 km intervals vertically for the accuracy of line synthesis.

\subsection{RH1.5D Synthesis}
We used the RH1.5D code \citep{2015Pereira} to synthesize the line profile. The Fe I 6569.2 \AA\ data are obtained from the Kurucz line database \citep{2018Kurucz}. For each column along the vertical direction, the radiative transfer function is calculated assuming local thermodynamic equilibrium (LTE) and complete redistribution (CRD). Compared with more realistic 3D Non-LTE radiative transfer, this column-by-column approach introduces only minor differences (see \cite{2015Pereira} and references therein). A critical consideration is the blending of the Fe I line with the far red wings of H$\alpha$. We find that the Fe I line profile remains invariant whether H$\alpha$ is treated in LTE or non-LTE. Consequently, we always use LTE approximation in calculations to achieve great computational savings. The magnetic field in the simulation does not lead to an obvious difference in the test, so it is set to 0. The microturbulent velocity is set to about 4 km/s to match the width of the observed profile at the disk center.

To mimic the image observed from the solar limb with a heliocentric angle $\theta$, we assume that the line-of-sight and the z direction of the simulation cube have an angle of $\theta$, and the projection of the line-of-sight is along the y axis. The properties of the simulation data are then interpolated to this new slanted direction. The line-of-sight velocity can be calculated as $v_\mathrm{los} = v_\mathrm{z}\cos\theta - v_\mathrm{y}\mathrm{sin}\theta$. The y-axis scale in the projected image is compressed by a factor of $\mu$ to account for the projection effect. Similar to the observation, we sample $\mu$ from 0.1 to 1.0 in steps of 0.1, generating ten distinct images corresponding to different limb positions.

\section{Result} \label{Result}

\begin{figure*}[htbp!]
\centering
\includegraphics[width=\textwidth]{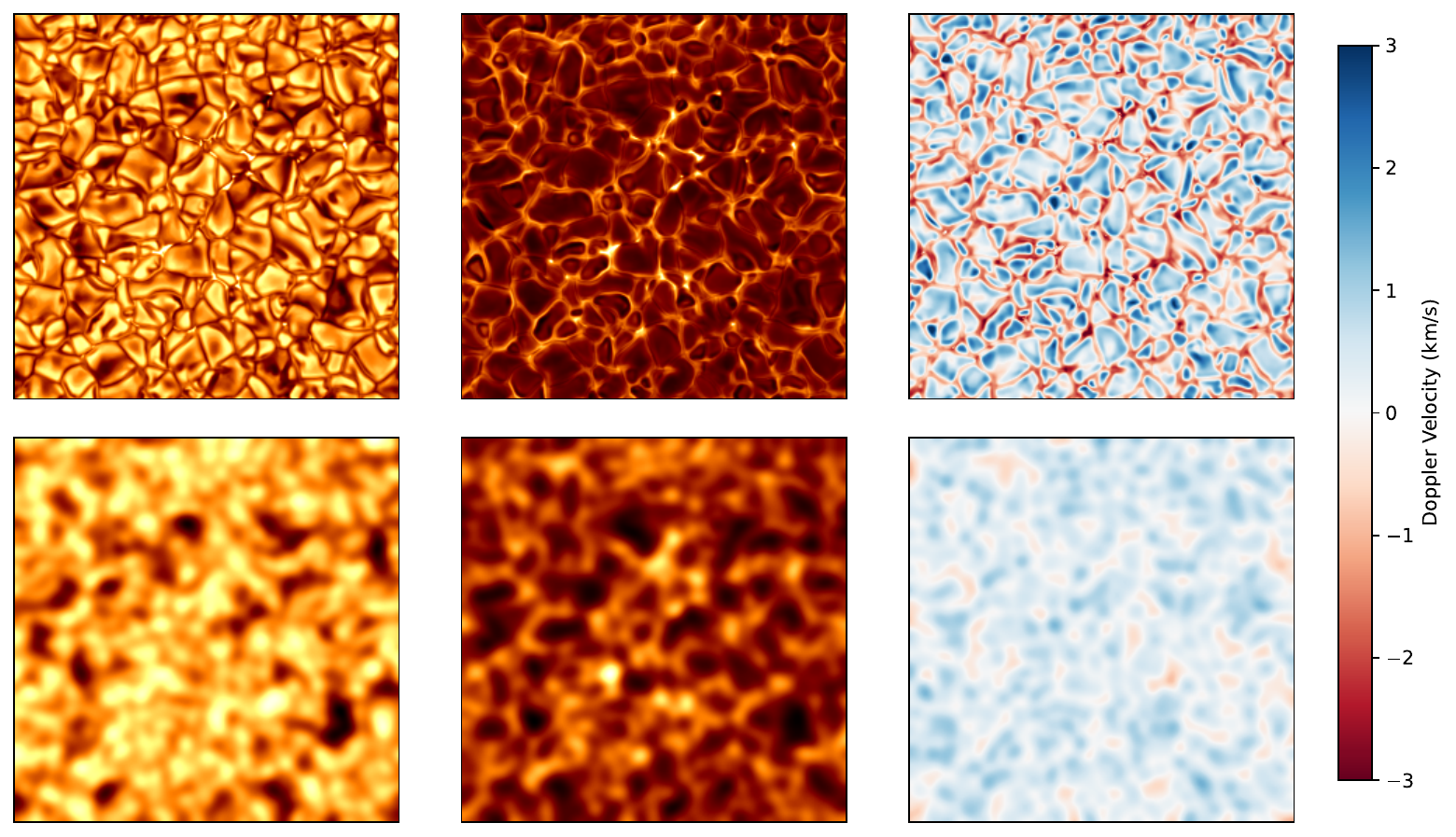}
\caption{Left: Synthetic Fe intensity images in the far wing (+0.78 \AA). Middle: Corresponding images in the line center. Right: Derived Dopplergrams and their colorbars. The images are shown with native simulation spatial resolution (top) and with the CHASE spatial resolution (bottom).}
\label{image}
\end{figure*}

Synthetic images on the solar disk center are shown in Figure~\ref{image}. For the spectrum of each pixel, we calculate the Doppler velocity using the same bisector method as the observation. It is clear that the upward-moving granule is surrounded by the intergranular network that moves downward. The granule is brighter than the intergranular lane in the line wings, corresponding with features observed in the solar continuum. In the line center, the reversed granulation pattern is found, with a darker center and a brighter edge. The phenomenon is caused by the imbalance between radiative heating and cooling of fluid elements \citep{2007Cheung}. To enable direct comparison with the CHASE observations, we degrade the images at each wavelength by convolving them with a Gaussian point spread function (PSF) of $1.2^{\prime\prime}$ FWHM, matching the telescope's spatial resolution. After convolving, the features of granules and reversed granules can be partially found in the intensity images. The Dopplergram exhibits some of the granular signature, but the redshift in the network basically disappears, and the blueshift part dominates across the entire field of view. This suggests that Method 2 is inherently sensitive to spatial resolution, which is discussed in Sect~\ref{dependence on spatial resolution}.

\subsection{Convective Blueshift}
\label{Calibration}
In the CHASE wavelength calibration pipeline, the mean Dopplergram at disk center is assumed to be zero after correcting for known shifts (e.g., fake Doppler velocity and gravitational redshift). However, this assumption introduces a systematic bias because it neglects the intrinsic convective blueshift inherent to photospheric absorption lines. By comparing with our simulation results, we quantify this bias and apply a blueshift offset to account for the convective motions.

\begin{figure*}[ht!]
\includegraphics[width=\textwidth]{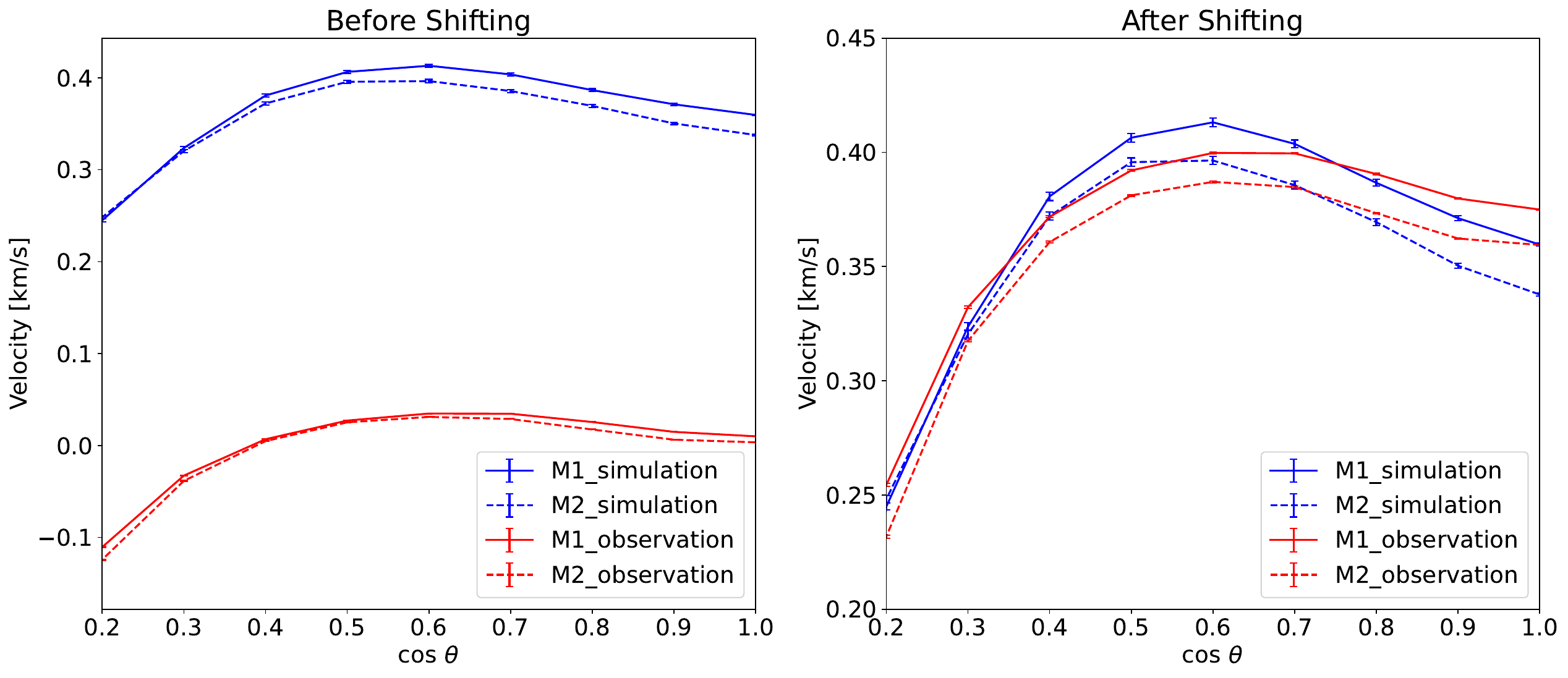}
\caption{Limb shift curves and their error bars from simulation and observation derived from Method 1 and Method 2, which are shown both with (left) and without calibration (right).}
\label{comparison}
\end{figure*}

For each heliocentric angle, we first degrade the images to CHASE's spatial resolution, then calculate the Doppler velocity using both Method 1 and Method 2. As shown in Fig.~\ref{comparison}, the simulated limb shift curves from both methods exhibit small differences and reproduce the observed trend: increasing blueshift from disk center to intermediate latitudes ($\mu \approx 0.6$), followed by a decrease toward the limb. The simulations display a systematic blueshift offset of about 0.35 km/s relative to observations. This discrepancy suggests a velocity component $V_c$ compared with the static wavelength. We therefore apply a least-squares optimization to determine the optimal $V_c$ that minimizes the residuals between simulated and observed curves. The derived offsets are $V_c = 0.366$ km/s in Method 1 and 0.356 km/s in Method 2. After applying this shifting, the adjusted limb shift curves (right panel of Fig.~\ref{comparison}) show an agreement with simulations, with deviations below 0.03 km/s across all heliocentric angles.

\subsection{Dependence on spatial resolution} \label{dependence on spatial resolution}
\begin{figure*}[ht!]
\includegraphics[width=\textwidth]{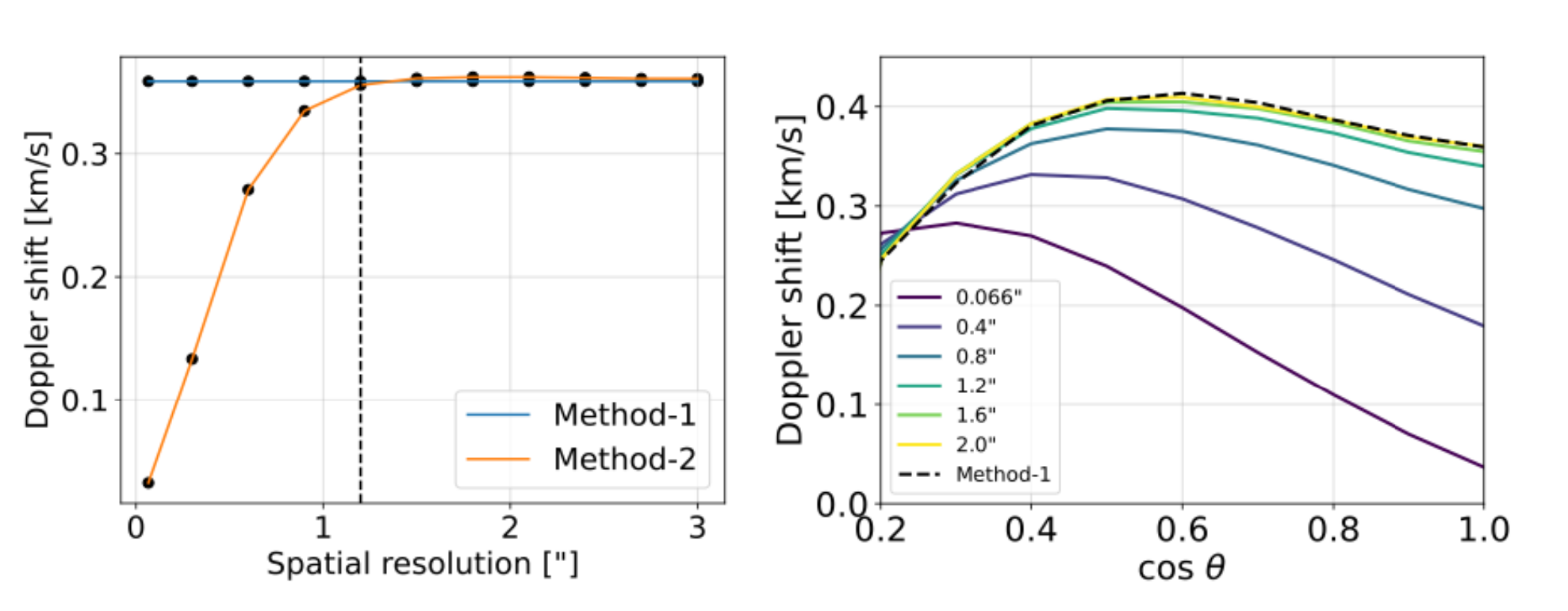}
\caption{Left: Velocity at the disk center derived from Method 1 (blue) and Method 2 (orange); vertical dashed line denotes the CHASE's spatial resolution ($1.2^{\prime\prime}$). Right: Limb shift curves of Method 1 (black dashed line) and Method 2 (colored solid line).}
\label{resolution}
\end{figure*}

We now explore how the limb shift curve is affected by spatial resolution. To simulate real observations, we vary the FWHM of the PSF when convolving the images. The results are exhibited in Fig~\ref{resolution}. The native simulation spatial resolution without convolution is significantly higher ($0.066^{\prime\prime}$). At the disk center, the velocity derived from Method 1 remains constant at 0.36 km/s, independent of spatial resolution. This is because convolution with a normalized PSF does not change the mean value at each wavelength, leaving the averaged spectrum unchanged. However, the velocity derived from Method 2 is only 0.06 km/s at simulated spatial resolution. As spatial resolution degrades (i.e., as the PSF width increases), the Method 2 results get closer to those of Method 1.

The right panel in Fig~\ref{resolution} shows the limb shift curves derived from Method 2 at different spatial resolutions. The dependence of spatial resolution is most pronounced near the disk center. At native resolution, Method 2 produces an anomalous trend in which the inferred blueshift increases from $\mu$ = 1.0 to 0.3 before decreasing, which is apparently different from the observed trend. It indicates that when using a high-resolution Dopplergram to measure the limb shift, the limb region ($\mu = 0.2$ - $0.3$) appears blueshifted relative to the disk center. Such a trend is contrary to the result of all previous measurements. At the extreme limb ($\mu \sim 0.1$), the shift becomes red again, although this behavior is not shown here. 

As spatial resolution degrades, the Method 2 curve progressively converges toward the Method 1 results. When spatial resolution is larger than $2.0^{\prime\prime}$, the two methods yield nearly identical profiles. It suggests that spatial smoothing diminishes the discrepancies between the methods: the PSF convolution spatially averages the surrounding signal onto each pixel. Consequently, the derived Dopplergram reflects the mean spectral profile over the convolved region, thus rendering Method 2 effectively equivalent to Method 1.

\subsection{Dependence on Spectral resolution} \label{dependence on spectral resolution}

\begin{figure*}[ht!]
\includegraphics[width=\textwidth]{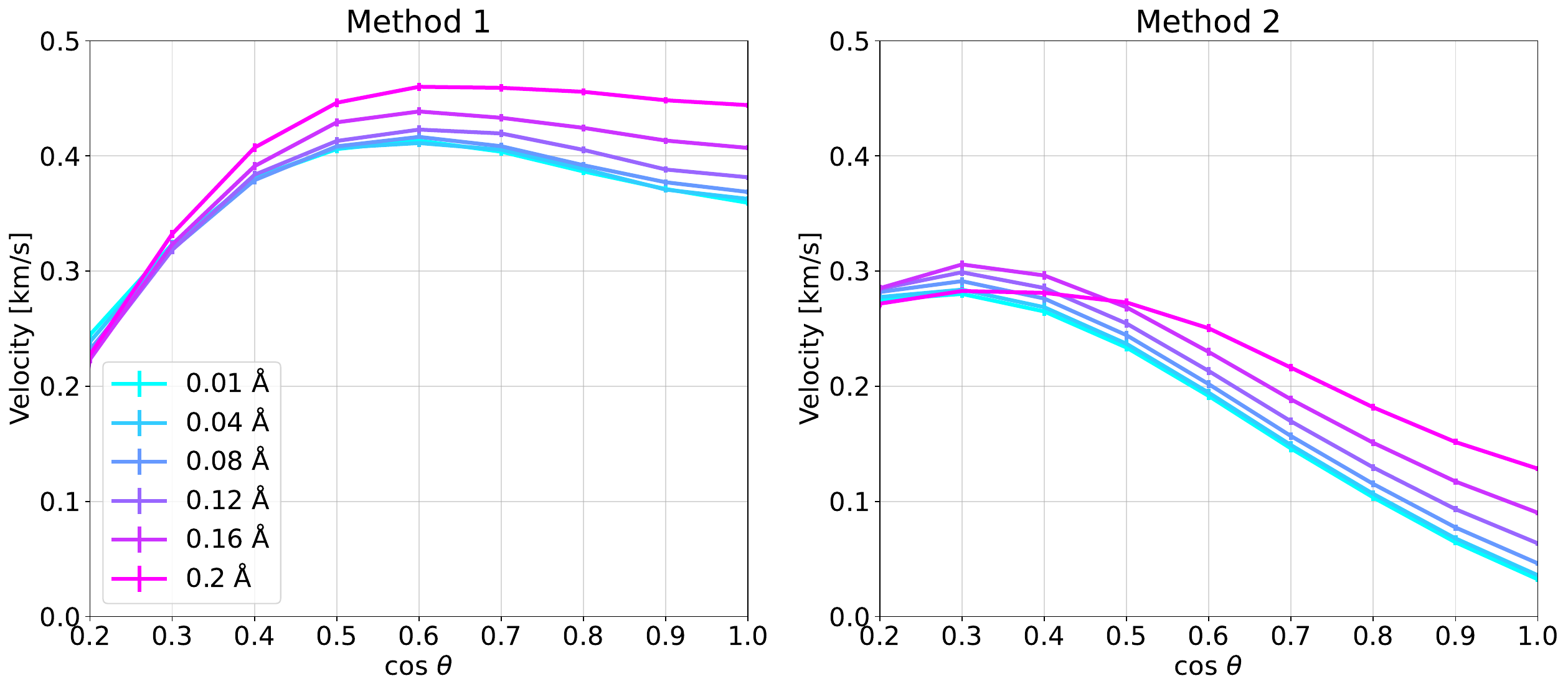}
\caption{Limb shift curves in the simulation, each synthetic spectral line is degraded into a lower spectral resolution.}
\label{spectral_resolution}
\end{figure*}

Using simulation data, we present the influence of spectral resolution on the limb shift in Fig.~\ref{spectral_resolution}, with the spatial resolution fixed at the simulated value. Each spectral profile was degraded to a lower resolution by convolution with a Gaussian kernel. For Method 1, as the spectral resolution decreases, the inferred blueshift increases at the disk center but decreases slightly near the limb ($\mu = 0.2$). The result in the disk center is consistent with the findings of \cite{2011Rodriguez} and \cite{2018Bottcher}. Regarding Method 2, the blueshift measured at disk center is also larger at lower spectral resolutions; towards the limb, such a correlation is not obvious. In general, the limb shift does not change significantly when the spectral resolution is better than 0.016 \AA\ ($R > 400000$ in the visible range).

\section{Origin of limb shift}
\label{Origin}

\begin{figure}[ht!]
\includegraphics[width=\linewidth]{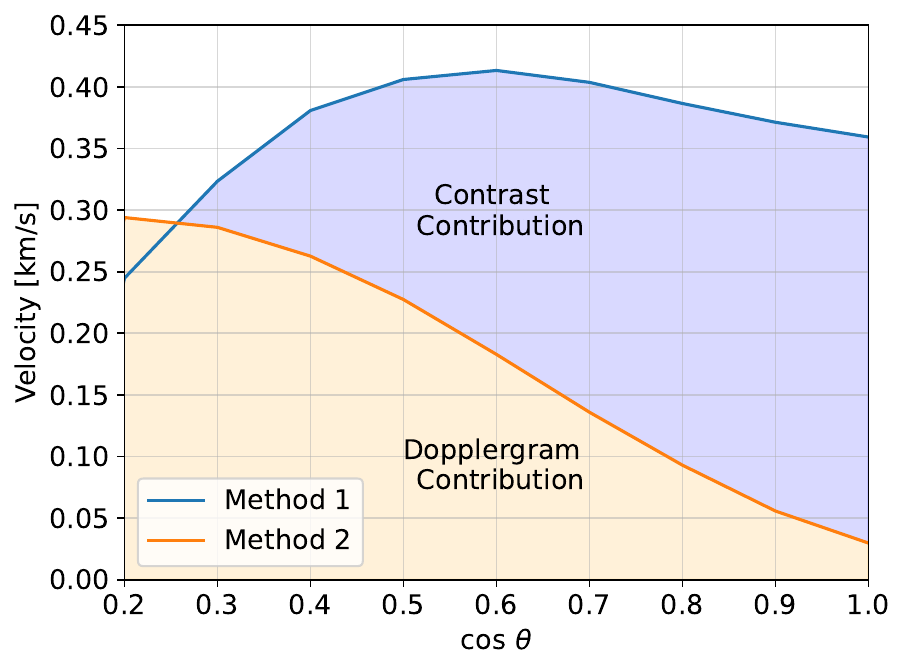}
\caption{Limb shift curves from Method 1 (blue) and Method 2 (orange) with the simulation's spatial resolution. The blue and orange shadows correspond to contrast and Dopplergram contributions, respectively.}
\label{contribution}
\end{figure}

In the idealized limit of perfect spatial resolution, the limb shift curves from Method 1 and Method 2 are displayed in Fig~\ref{contribution}. The origin of limb shift can be understood as follows. 

In Method 2, the Doppler shift is computed for each pixel before any averaging is performed. As a result, all pixels are weighted equally, regardless of whether they correspond to bright granules or darker network regions. Even under equal weighting, the spatially resolved Dopplergram exhibits a net blueshift. This intrinsic component — hereafter referred to as the Dopplergram contribution — is illustrated by the orange shading in Fig. \ref{contribution}.
    
In contrast, Method 1 first averages the spectra, so granules and intergranular lanes no longer contribute equally; this unequal weighting introduces an additional blueshift on top of the Dopplergram contribution. The difference between Method 1 and Method 2 then represents the contrast contribution, illustrated by the blue shading in Fig.~\ref{contribution}.

From disk center to limb, the Dopplergram contribution becomes increasingly dominant, while the contrast contribution weakens and even becomes negative at $\mu = 0.2$. This behavior will be further analyzed using simulated data in the following sections.

\subsection{Contrast Contribution}\label{contrast contribution}

The contrast contribution, often referred to as the correlation between Doppler shifts and line intensity, is commonly regarded as the main cause of the convective blueshift \cite{1978Beckers(b),2000Asplund,2011Rodriguez,2021Kashyap}.
In the classical picture \cite{1981Dravins}, bright upflowing granules and dark downflowing intergranular lanes together yield an averaged, blueshifted line profile.

\begin{figure*}[htbp!]
\centering
\includegraphics[width=\textwidth]{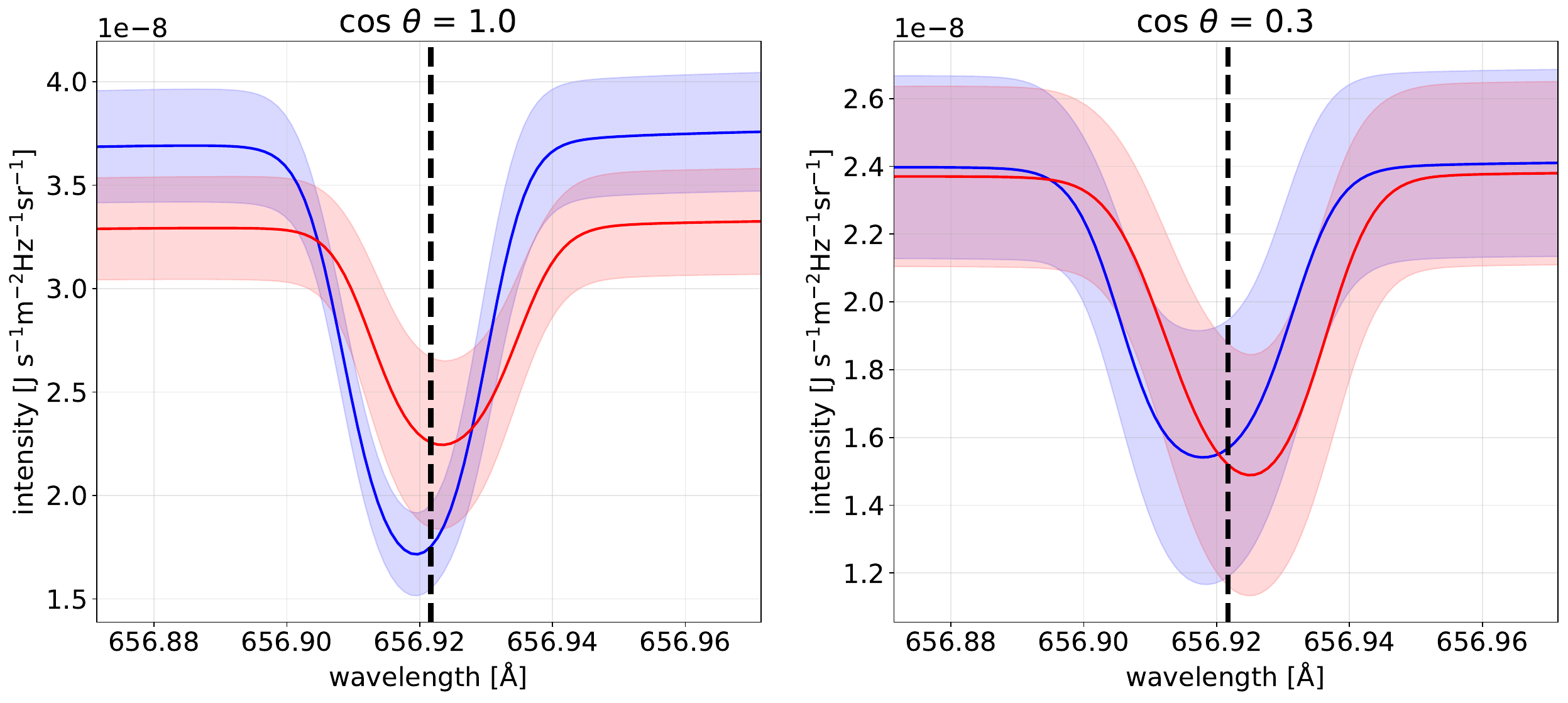}
\caption{Averaged red/blue shifted spectrum across the simulated image at $\mu$ = 1.0 (left) and 0.3 (right). Shaded areas show the standard deviation at different wavelengths. Vertical dashed line denotes the line center of each panel.}
\label{bright}
\end{figure*}

To illustrate this effect, we classify the pixels in our synthetic data into blue-shifted (refer to granule) and red-shifted (refer to intergranular lane) groups and compute their average profiles, as shown in Fig~\ref{bright}. At the disk center, blue-shifted granules are not always brighter across wavelength. It is obvious that granules are brighter than intergranular lanes in the continuum (normal granule pattern), but darker in the line center (reversal granule pattern). Compared with the classical description, the behavior in our simulation is therefore more complex. Therefore, we describe spectral-line profiles in granules as being ``deeper'' than intergranular lanes rather than simply ``brighter''. And this depth contrast is what contributes to the net blueshift at disk center. When $\mu = 0.3$ (right panel in Fig~\ref{bright}), the red and blueshift profiles have nearly the same intensity and depth, and even red-shifted profiles are deeper. As a result, the contrast contribution gradually disappears towards the limb and even becomes reversal at $\mu = 0.2$ in Fig~\ref{contribution}.

\subsection{Dopplergram Contribution}
The mean Dopplergram shows a net blueshift even if there is no contrast contribution. It is related to the velocity where the line is formed and the area of the blue/red-shifted region. We further divide this contribution into two components.

\subsubsection{Inhomogeneous Density}\label{Inhomogeneous Density}
The selected simulation data at t = 5280 s has nearly zero net average mass flux along the line-of-sight in the line formation region, consistent with the limb shift condition defined by \cite{1988Ulrich}. This condition requires $\int \rho v_\mathrm{L} \mathrm{dS}=0$, where $\rho$ is the mass density and $v_\mathrm{L}$ is the line-of-sight velocity. However, inhomogeneities in the density field (e.g., $\rho /v_\mathrm{L}$ correlation) will cause $\int v_\mathrm{L} \mathrm{dS}>0$, as demonstrated by \cite{2016Carlsson}. Our calculations of the mean velocity across different atmospheric layers reveal a persistent redshift of approximately $100$ m/s in the line formation region (located at 150 km for $\mu$ = 1.0 and 300 km for $\mu$ = 0.1 in our model). This implies that density inhomogeneities introduce a redshift contribution of $\sim100$ m/s to the limb shift, independent of the heliocentric angle. We note that this effect may rely on the different models of the atmosphere.

\subsubsection{Corrugation effect}
In Sect.~\ref{Inhomogeneous Density}, we assume that the mean Dopplergram arises from a geometrically flat layer. However, the line-of-sight conditions vary from column to column, causing the spectral line at different spatial positions to sample information from different depths in the photosphere. Therefore, the line formation surface exhibits corrugated rather than flat layer\citep{1998Stein}. This corrugation also accounts for part of the Dopplergram contribution, while it presents a larger effective area of blueshifted regions than redshifted regions, especially at the limb.

\begin{figure}[ht!]
\includegraphics[width=\linewidth]{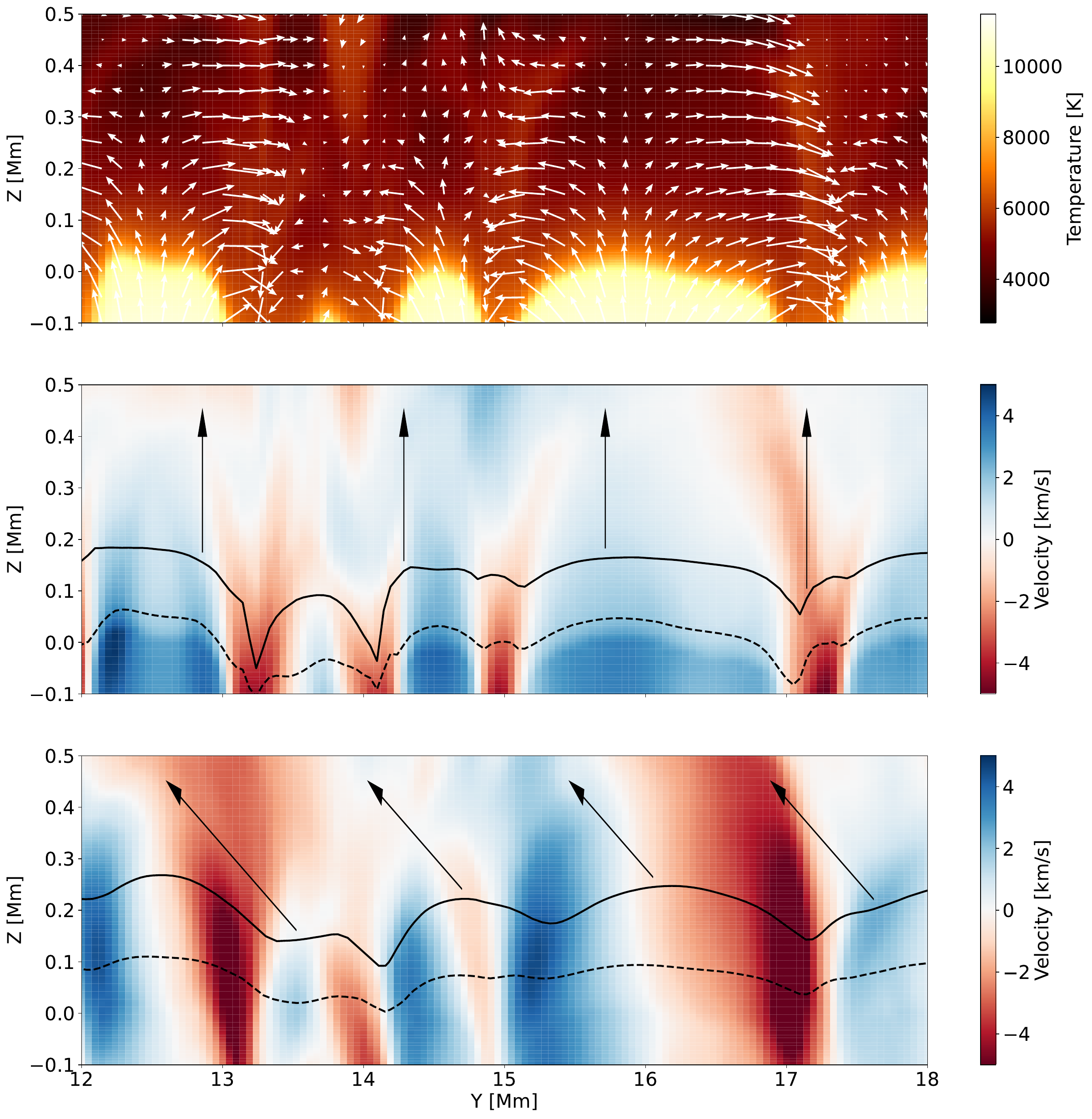}
\caption{Top: Temperature distribution in the selected y-o-z plane slice. White arrows indicate the bulk velocity field. Middle: Line-of-sight velocity in the same slice when $\mu=1.0$. Black arrows mark the viewing direction. Solid and dashed line denotes the $\tau=1$ layer in the line center and in the 500 nm continuum, respectively. Bottom: Same as the middle panel but for $\mu=0.3$}
\label{corrugation}
\end{figure}

Following \cite{2011Rodriguez}, we show structures of the temperature and the line-of-sight velocity in the y-o-z plane slice in Fig~\ref{corrugation}. The $\tau=1$ layer exhibits significant corrugation both at $\mu=1.0$ and 0.3, with peaks at the granules and valleys aligned with intergranular lanes. When we look in a tilted direction ($\mu=0.3$), the horizontal velocity is projected onto the line-of-sight, causing the left part of the granule to contribute to the blueshift while the right part contributes to the redshift. This effect is obvious in granules located at $Y=12\sim13$ Mm and $Y=15\sim17$ Mm, and is common in other slices. In this way, the $\tau=1$ layer in the redshift part is more parallel to the line-of-sight, so it shows relatively smaller apparent area to the observer than the blueshift part. Then it manifests as a mean blueshift. Such a projection effect is caused by the horizontal velocity; therefore, it does not exist at the disk center and becomes more pronounced towards the limb because of the higher proportion of the horizontal velocity in the line-of-sight direction.

\cite{1985Balthasar} have analysed this phenomenon and find it happens at $0.6 < \mu < 1.0$ and becomes reversed at the disk limb due to the reversed granulation pattern (the $\tau=1$ layer exhibits valleys at the granules and peaks at the intergranular lanes). However, our model shows no such reversal at $\mu=0.3$, since the $\tau=1$ layer retains a shape similar to that at $\mu=1.0$. This persistence confirms that corrugation consistently contributes to blueshifts across all viewing angles in our simulations.

It should be mentioned that the median of the velocity response function, where the velocity samples are the better choice than the $\tau=1$ height \citep{1969Beckers,1975Beckers}. The discussion based on the $\tau=1$ layer introduces systematic errors. These limitations suggest that future work should incorporate response function-weighted averaging to improve accuracy.

\section{Discussion}\label{Discussion}

\subsection{Comparing with previous results}

\begin{figure}[ht!]
\includegraphics[width=\linewidth]{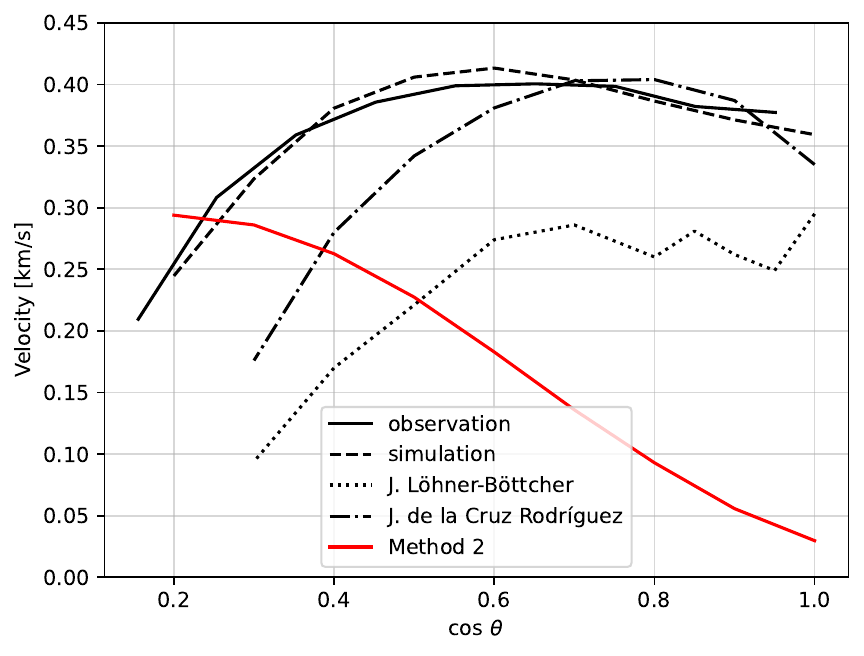}
\caption{Limb shift in the CHASE observation and Bifrost simulation, in comparison with previous results. The red line shows the simulation results from Method 2, with the simulation’s spatial resolution.}
\label{previous}
\end{figure}

We plot the previous limb shift result (\cite{2011Rodriguez} and \cite{2018Bottcher}) of Method 1 in comparison with our observation and simulation in Fig~\ref{previous}. It should be noted that the lines analyzed by \cite{2011Rodriguez} and \cite{2018Bottcher} (Fe I 6302.49 \AA) form at different atmospheric heights than the Fe I 6569.21 \AA\ line used in this study, which may naturally lead to systematic differences.

Each curve is derived from Method 1, except for the red line, which represents Method 2. At the disk center, Method 1 introduces an average blueshift of approximately 0.35 km/s for all results, which is consistent with the traditional values \citep{1981Dravins}. The limb curve of \cite{2011Rodriguez} peaks around $\mu = 0.8$ at 0.4 km/s before decreasing rapidly. In our work, the limb shift curve peaks at $\mu = 0.65$ in observation and 0.6 in simulation. Notably, our results decrease much more gradually towards the limb, with a rapid decline only occurring at $\mu < 0.3$. For instance, at $\mu = 0.3$, the blue shift is only 0.1 km/s in \cite{2018Bottcher} but 0.35 km/s in our work. We attribute this systematic discrepancy primarily to the different lines and different velocity measurements employed: both \cite{2011Rodriguez} and \cite{2018Bottcher} use the mean bisector to measure the velocity, while we use the single bisector at the 0.7 continuum intensity, which is known to yield the maximum blueshift magnitudes. Such an effect may contribute to the discrepancy.

The red line shows the result from Method 2 at high spatial resolution, which is already displayed in Fig~\ref{resolution}. The velocity monotonically increases from 0.03 km/s at the disk center towards the limb, and reaches 0.3 km/s at the limb. Such a trend is totally in contrast with other curves. As mentioned in Sec~\ref{Origin}, Method 2 does not contain brightness information, so the differences are mainly caused by the contrast contribution. This suggests that such a trend should become observable when instruments reach sufficiently high spatial resolutions.

\subsection{Application of two methods}

Following the definition of the limb shift curve \citep{1976Beckers}, we should use Method 1 in the observation. Considering that Method 2 is used in some of the works \cite{2010Ulrich,2021Kashyap}. It is necessary to distinguish the application of the two methods.

In Sect~\ref{dependence on spatial resolution}, we mention that the limb shift from two methods has almost no difference when the instrument has a low spatial resolution ($>2.0^{\prime\prime}$). In this case, both methods are acceptable as they finally lead to the same results. 

For an instrument with high spatial resolution, different methods introduce significantly different limb shift curves (Fig~\ref{resolution}). At the disk center, the average blueshift of Method 2 is much smaller than Method 1. In this situation, the choice of method depends on the research objective. For example, the purpose of \cite{1988Ulrich,2024Rao} is to calculate the differential rotation, while the limb shift is fitted from the observation and eliminated from the Dopplergram. Then, we should use Method 2, even though the obtained curve may differ from the traditional one. For observations that aim to measure limb shift \citep{2011Rodriguez,2018Bottcher,2018Cegla}, Method 1 is a better choice as the definition. However, the spatial resolution of these observations is not high enough to introduce an obvious difference between the two results.

\begin{figure}[ht!]
\includegraphics[width=\linewidth]{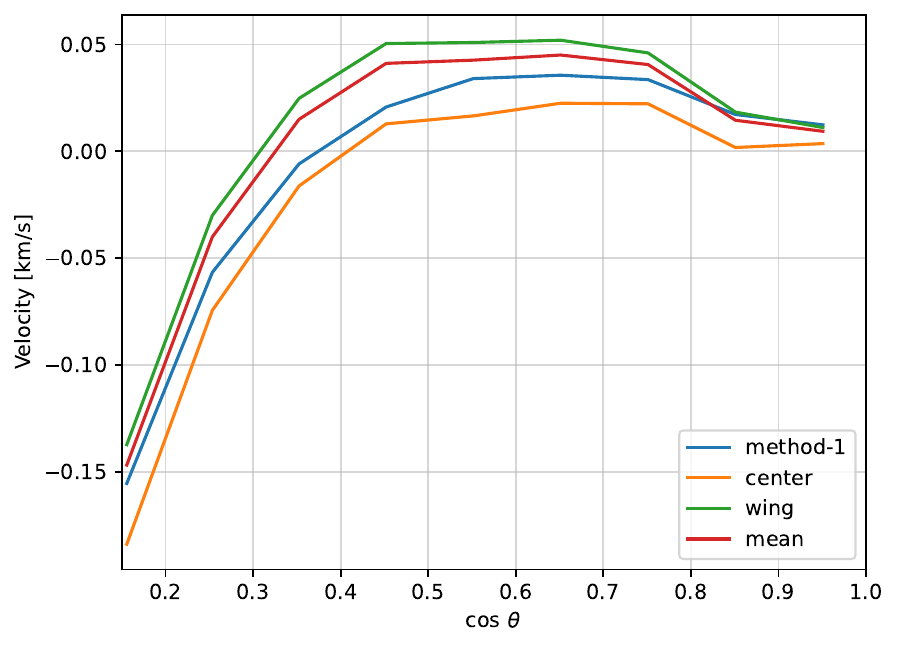}
\caption{Limb shift results calculated by Equation~\ref{third method} in the CHASE observation, but with $I_{ij}$ in the line center (orange), line wing (green), and mean intensity across the wavelength (red). Blue line displays the result of Method 1.}

\label{method3}
\end{figure}

In addition to these two methods, \citep{2024Palumbo} applies another way to measure the limb shift curve. They calculated the average velocity weighted by intensity for each heliocentric angle,
\begin{equation}
    \bar{v} = \frac{\sum_{ij} v_{ij}I_{ij}}{\sum_{ij} I_{ij}}.
    \label{third method}
\end{equation}
Here, the intensity information $I_{ij}$ is added in the Dopplergram $v_{ij}$, making the result similar to Method 1 \citep{2018Cegla}. However, the intensity must be carefully considered. As we discussed in Sec~\ref{contrast contribution}, the upward-moving granules are brighter than intergranular lanes in the line wings, but darker in the line center due to the reversed granulation. The inferred averaged blue shift will become smaller when weighted by the line core intensity and larger when weighted by the line wing intensity. We show the observation result by this method in Fig~\ref{method3}. Consequently, caution is advised when applying this method to lines that exhibit a pronounced reversed granulation pattern in observations.

\section{Conclusion}
\label{conclusion}
In this work, we analyse the limb shift effect of the line Fe I 6569 \AA\ using the CHASE observation. The limb shift curves are derived through the spectral-averaging method and the velocity-averaging method. Two methods show similar results. To validate these findings, we synthesize the same spectral line from Bifrost simulation data and apply both methods. The conclusions are summarized as follows.

\begin{enumerate}
    \item At the CHASE spatial resolution, the simulations reproduce curves that match the observations from both Method 1 and Method 2. These results allow us to determine a systematic blueshift in the CHASE wavelength calibration, which is not accounted for in the current pipeline. The convective motions introduce a blueshift of approximately 0.36 km/s.
    
    \item  At high spatial resolution ($<1^{\prime\prime}$), the limb shift curve from Method 2 shows strong departurefrom classical limb shift profiles \citep{1960Higgs,1988Ulrich,2018Bottcher} (Fig.~\ref{resolution}). As the resolution descends, the curve from Method 2 converges toward that from Method 1, suggesting resolution-dependent systematic effects. It underlines the necessity of method selection based on instrumental resolution.
   
    \item At high spatial resolution, the limb shift result consists of two distinct components (Fig.~\ref{contribution}): a contrast contribution and a Dopplergram contribution. The contrast contribution arises from the unequal weight of granules and intergranular lanes, which decreases toward the limb. In contrast, the Dopplergram contribution, driven by density inhomogeneities and corrugation effects in the $\tau=1$ layer, increases dominantly toward the limb. These opposing effects explain the origin of the limb shift curve (Fig~\ref{contribution}).

\end{enumerate}

We note that the contrast and Dopplergram contributions discussed above are only effective at sufficiently high spatial resolution. As the spatial resolution decreases, the Dopplergram contribution gradually becomes blended with the contrast contribution due to the convolution effect (see the intermediate curves in the right panel of Fig.~\ref{resolution}), while the sum of these two remains unchanged. Importantly, this requirement does not impose extremely stringent demands on spectral resolution (refer to Fig.~\ref{spectral_resolution}). Modern high-resolution facilities—such as DKIST, SST/CRISP, Sunrise/SCIP, and BBSO/GST already offer sufficient spatial resolving power to perform such tests. To further generalize these results, a systematic investigation of additional photospheric spectral lines will help assess.

\begin{acknowledgements}
This project has been funded by National Key R\&D Program of China under grants 2021YFA1600504 and 2022YFF0503004, and by NSFC under grants 12127901 and 12333009. J.H. is funded by the European Union through the European Research Council
(ERC) under the Horizon Europe program (MAGHEAT, grant agreement 101088184). The Institute for Solar Physics is supported by a grant for research infrastructures of national importance from the Swedish Research Council (registration number 2021-00169).
 \end{acknowledgements}

\bibliography{sample701}{}
\bibliographystyle{aasjournalv7}

\appendix

\section{Observations in hemisphere} 

    \begin{figure}[h]
    \centering
    \includegraphics[width=\linewidth]{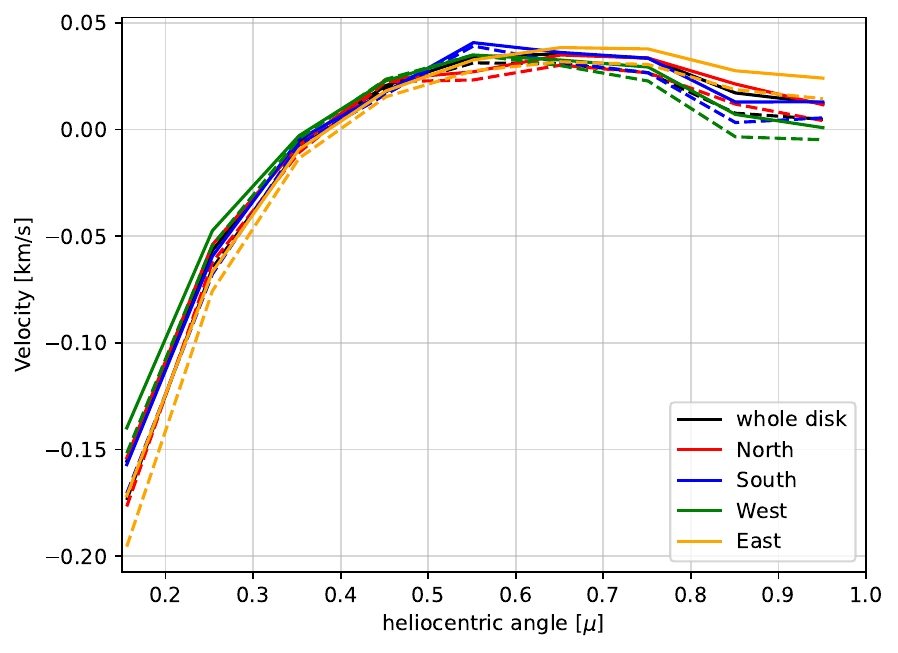}
    \caption{Limb shift curves using data of North (red), South (blue), West (green), and East (orange) hemispheres. Solid and dashed lines denote Method 1 and Method 2, respectively.}
    \label{hemisphere}
    \end{figure}

\end{document}